# Experimental studies on vacancy induced ferromagnetism in undoped $TiO_2$

Abdul K Rumaiz[1], Bakhtyar Ali[1], Abdullah Ceylan[1, 4], M. Boggs[2], T. Beebe[2] and S. Ismat Shah[1, 3]

1. Department of Physics and Astronomy
2. Department of Chemistry and Biochemistry
3. Department of Materials Science and Engineering

University of Delaware, Newark DE 19716

4. Physics Engineering Department, Hacettepe University, Ankara, Turkey.

**Abstract**

Room temperature ferromagnetism is observed in undoped $TiO_2$ films deposited on Si substrates using pulsed laser deposition (PLD). The ferromagnetic properties of the samples depend on the oxygen partial pressure during the PLD synthesis. The appearance of higher binding energy component (HBEC) in the oxygen 1*s* core peak from x-ray photoelectron spectroscopy (XPS) suggests the presence of oxygen vacancies in these samples. The amount of oxygen during the synthesis determines the vacancy concentration in the samples which is directly related to the magnetic behavior of the samples. The magnetic moment decreases with oxygen vacancy concentration in the samples. Valence band measurements were performed to study the electronic structure of both stoichiometric and reduced $TiO_2$. The analyses show the presence of Ti 3*d* band near the Fermi level in reduced $TiO_2$ samples. These bands are otherwise empty in stoichiometric $TiO_2$ and reside in the conduction band which makes them unobservable by XPS. The existence of this Ti 3d band near the Fermi level can possibly lead to Stoner splitting of the band.

## Introduction

Transition metal oxides have been extensively studied because of their technological importance. For example, the usefulness of $TiO_2$ for photocatalytic applications has been well known [1, 2]. Following the discovery of room temperature ferromagnetism in transitional metal (TM) doped oxide semiconductors [3], several studies have been carried out to understand the origin of ferromagnetism in these systems. There is an ambiguity about the origin of the ferromagnetism. However, some recent results have shed light on the intrinsic nature of ferromagnetism in these systems [3-5]. The quest for finding a room temperature dilute magnetic semiconductor (DMS) is especially crucial in the field of spintronics where the spin property of the carrier along with its charge property is utilized. Apart from TM doped oxide semiconductors, various other DMS such as Mn doped II-VI and III-V compound semiconductors have been extensively studied [7,8]. In most of these materials the Curie temperature is much lower than room temperature. On the other hand, oxide semiconductors have Curie temperature higher than the room temperature, have a large band gap and can easily be doped with n-type dopants. Recently, Coey's group reported magnetism in pure HfO films [9]. Following this, Hong et. al. [10] and Yoon et. al. [11] reported room temperature ferromagnetism in pure $TiO_2$ films. This is a surprising result and has generated a lot of interest in this new phenomenon, also known as $d^0$ magnetism [12]. In these systems it is known that oxygen vacancies behave as an n-type dopant, and the existing vacancies might be the reason for magnetic order. Also, in the electronic structure calculation done by Paxton and Thien-Nga [13], it is shown that when the Ti derived conduction band states are occupied upon reduction, there exists a possibility of magnetic ordering due to states appearing in the band gap that have split-off from the conduction band.

Oxygen vacancies have a profound influence on the physical and chemical properties of transition metal oxides [14] in which they act like electron donors [15]. It is well known that $TiO_2$ is very sensitive to oxygen and can be very easily reduced under an oxygen deficient environment. For the doped $TiO_2$ there is an additional charge neutrality condition that has to be satisfied when the dopants are non-isovalent. This charge neutrality condition leads to the formation of Ti or O vacancy, depending on the valency of the dopant. In fact, some authors give this argument to explain the magnetic and transport behavior of Co doped $TiO_2$ [16]. There have also been some reports which had linked oxygen vacancies with ferromagnetism in TM doped ZnO and HfO samples [17,18].

Although the recent report from Hong et al showed ferromagnetism in pure $TiO_2$ samples, in our present work we studied the effect by systematically changing the oxygen vacancy concentration in our samples. This was achieved by changing the oxygen partial pressure during the synthesis of the samples by pulsed laser deposition (PLD). We found that the films are ferromagnetic in a very small range of oxygen partial pressure. Ferromagnetism vanished for sample prepared in high oxygen partial pressure. The electronic structures of both stoichiometric and reduced $TiO_2$ were also studied.

## Experimental

The $TiO_2$ films used in this study were prepared by pulsed laser deposition technique. The undoped solid $TiO_2$ target was prepared by pressing highly pure (99.999% Sigma-Aldrich) titanium dioxide powder into a 5cm diameter and 0.5 cm thick disk . The targets were ablated using an excimer laser (Lambda Physik LPX 305, λ=248 nm) at a constant fluence of 1.8 J/cm$^2$. The films were deposited on Si and quartz substrates at 800°C. The base pressure in the chamber was ~10$^{-7}$ Torr. Immediately prior to the deposition, the chamber was backfilled with a high purity (99.9%) oxygen gas. Samples were prepared under several different oxygen pressures of 0.02 mTorr, 0.2 mTorr, 2mTorr, 50 mTorr and with no oxygen introduced in the chamber, hereafter referred to as 0 mTorr sample. Care was taken to handle all samples with Teflon tweezers to avoid magnetic contamination [19]. The structural characterizations were performed by X-ray diffraction (Rigaku Dmax B) using Cu Kα radiation. All magnetic measurements were carried out in a Quantum Design Magnetic Property Measurement System (MPMS) SQUID magnetometer. XPS data were obtained using a VG ESCALAB 220i-XL electron spectrometer (VG scientific Ltd., East Grinstead, U.K). Monochromatic $AlK_\alpha$ X-rays (1486.7 eV) were employed. Typical operation conditions for the X-ray source were a 400 $\mu$m nominal X-ray spot size (full width at half-maximum) operating at 15 kV, 8.9 mA, and 124 W for both survey and high-resolution spectra. Survey spectra from 0 to 1200 eV binding energy were collected at a 100 eV pass energy and a dwell time of 100 ms per point, and a total of two scans (averaged) in the respective binding energy ranges. High resolution XPS spectra were collected at 20 eV pass energy. The data acquisition was done using Eclipse data system software. The operating pressure of the spectrometer was typically $2x10^{-10}$ Torr The instrument was calibrated to Au 4*f* peak of a reference sample

## Results and Discussion

Figure 1 shows the XRD pattern for samples prepared under different partial pressures of oxygen. All films were polycrystalline. The peak positions for the samples prepared at the base pressures in the range of $10^{-7}$ Torr do not correspond to either anatase or rutile phases of $TiO_2$. At this low partial pressure of oxygen, the structure of $TiO_2$ becomes unstable and forms some other oxide [20]. The samples prepared at high oxygen pressures clearly show the dominant rutile (110) peak. From figure 1 it can be observed that for samples prepared at 0.02 mTorr, both rutile (110) and anatase (101) peaks are present. For a sample prepared at 50 mTorr, the XRD pattern contains predominantly anatase related peaks. Within the detection limit of XRD no trace of any metal impurities related peak was found in the XRD pattern.

$TiO_2$ films deposited at 0.02 and 0.2 mTorr partial pressures of oxygen are ferromagnetic at room temperature. Figure 2a shows the magnetization (M) versus field (H) curves taken at 298 K for the three samples prepared at 0.02, 0.2 Torr and 50 mTorr. Apart from the high field region that shows a negative slope because of the diamagnetic contribution from the substrate, the films prepared at 0.02 mTorr and 0.2 mTorr partial pressures are clearly ferromagnetic. The 0.02 sample shows small hysterisis with remnant magnetization of about 0.06 emu/cc and saturation magnetization of about 0.32 emu/cc. The samples prepared at 0.2 mTorr show significantly lower saturation magnetization, around 0.03 emu/cc. Since there is no other ferromagnetic dopant, these observations suggest that the origin of magnetism in these samples must be due to the oxygen vacancies that are created during synthesis. Samples prepared at 50 mTorr oxygen partial pressure shows diamagnetic behavior. The inset in figure 2a shows the magnetization versus field curve for the target material, which also clearly shows diamagnetic behavior. The large value of magnetic moment is hard to attribute to any kind of impurity. If such impurity did exist, we should have been able to observe it in the XPS spectra. Figure 3 shows the XPS survey scan of a film deposited at 0.02 mTorr oxygen partial pressure. The peaks observed above 900 eV correspond to Auger lines from $O_2$ and Ti. No impurity related peak could be observed. The substrates and the straw used to hold the sample during the magnetic measurement also showed a clear diamagnetic behavior. Figure 2b shows the temperature dependence of the magnetization of the 0.02 mTorr sample recorded with an applied field of 0.5 kOe, parallel to the film surface. The zero field and the field-cooled measurements do not show any spin-glass like behavior. The trend in the M-T curve can be attributed to the coexistence of the ferromagnetic and diamagnetic components, as reported in similar systems previously [21].

Due to the obvious difference in magnetic response related to the preparative conditions, we investigated the effect of oxygen vacancies in the XPS spectra. A 3 eV electron charg e neutralizer was used to neutralize surface charging. All the peaks reported were charge corrected using C 1*s* peak position at 284.6 eV as the reference point. Figure 4a shows the high-resolution O 1*s* XPS spectra of samples deposited at 0.02 mTorr and 50 mTorr gas pressure. The peak is deconvoluted into three symmetric peaks numbered 1-3. The peak at the highest binding energy (peak 3) can be attributed to the oxygen absorbed in water and the lowest binding energy (peak 1) is ascribed to the O 1*s* core peak of $O^{2-}$ bound to $Ti^{4+}$. There is an emergence of a new peak as the synthesis pressure decreases. This peak is referred to as the high binding energy component (HBEC) and the peak 1 is referred as the low binding energy component (LBEC) [22]. It has been previously reported that the HBEC component develops with the increasing loss of oxygen [23]. Figure 4b shows the relative area of the HBEC as a function of the oxygen partial pressure during deposition. From the trend it is obvious that we have more vacancies as the partial pressure of oxygen during deposition decreases. The fact that all the samples are prepared in identical condition also rules out the possibility that the growth of peak 2 is due to oxygen absorbed as water.

Stoichometric $TiO_2$ has $Ti^{4+}$ and is non-magnetic. However, unpaired 3d electron in $Ti^{3+}$ or $Ti^{2+}$ can lead to some magnetic moment. It is known that oxygen vacancies create charge imbalance and therefore there is a possibility to generate $Ti^{3+}$ and/or $Ti^{2+}$. The high-resolution Ti 2*p* spectra are shown in figure 5. It is clear that for the film deposited at 50 mTorr there are two peaks (referred to as peak 1 and peak 2). These are the characteristic $2p_{3/2}$ and $2p_{1/2}$ spin doublet from $Ti^{4+}$ at 458.3 and 464.1 eV, respectively, with a peak separation of 5.8 eV [24]. For the sample prepared at 0.02 mTorr, in addition to the above mentioned peaks, there are extra peaks labeled as peak 3 and peak 4. It might be argued that these two peaks correspond to $2p_{3/2}$ and $2p_{1/2}$ of $Ti^{3+}$ or $Ti^{2+}$. However, the reported value of the peak separation between $Ti^{4+}$ and $Ti^{3+}$ is around 1.8-1.9 eV [25]. The spectrum in fig 5 shows the peak separation to be on the order of 1.4-1.5 eV. The fact that the binding energy of peak 3 is in between the binding energies of $Ti^{4+}$ and $Ti^{3+}$ rules out the possibility of it being $Ti^{2+}$. This implies that the peak originates from core level peaks of $Ti^{4+}$ bound to an oxygen vacancy. We can equivalently consider peak 3 to have a slightly lower binding energy because the removal of oxygen will lead to a higher electron cloud density than for the ones corresponding to $Ti^{4+}$. This clearly rules out the formation of either $Ti^{3+}$ or $Ti^{2+}$. Thus both the features in the core level peaks of O 1*s* and Ti 2*p* strongly suggest the existence of oxygen vacancies in the system.

The valence band spectra for the reduced $TiO_2$ (0.02 mTorr) and stoichometric film prepared at high pressure (50 mTorr) are shown in figure 6. Pure $TiO_2$ has primarily a filled O 2*p* derived valence band separated from an empty Ti 3*d*, 4*s* and 4*p* derived conduction band by a bulk band gap of 3.2 eV [26]. The valence band spectra show the emission from O 2*p* band whose upper edge lies 3 eV from the Fermi level. The feature close to 0 eV in the reduced $TiO_2$ can be attributed to the occupied defect states corresponding to the partial population of the Ti 3*d* band [27]. This impurity level should grow with higher vacancy concentration. However, we could only see this band in samples prepared at 0.02 mTorr. This could be because the band position can also change such that it could possibly be below the Fermi level, in which case it would be difficult to see by XPS, as suggested by Henrich and Cox [27]. Figure 5 also shows that the O 2*p* valence band moves away from the Fermi level when vacancies are created. This can be understood by assuming a rigid band model with the oxygen vacancies moving the Fermi level to the conduction band; the occupied defect state would now correspond to the original empty conduction band states. Thus we see that in reduced $TiO_2$ the Fermi level falls on a region of high density of states. Furthermore, because the Ti 3*d* band is also narrow and the density of states is fairly large, we can expect a Stoner exchange splitting of the band which, in turn, might give rise to a magnetic moment on the Ti atom. This exchange splitting in the conduction band can be observed as a small chemical potential shift in the core level peaks [4, 28]. We have carefully measured the core level peak positions for three samples prepared at 0.02, 0.2 and 50 mTorr. Figure 7 shows Ti 2*p* and O 1*s* core level spectra for stoichiometric $TiO_2$ (prepared at 50 mTorr) and reduced $TiO_2$ (prepared at 0.2 and 0.02 mTorr). Although the nature of the peaks have been discussed earlier, we can see that the core peaks of both O *1s* and Ti *2p* are shifted by 0.5 and 0.3 eV for the samples prepared at 0.02 mTorr and 0.2 mTorr of oxygen partial pressure, respectively.

In conclusion, we have prepared undoped $TiO_2$ films and observed ferromagnetism in films deposited within a certain pressure range. The core level peaks of both Ti and $O_2$ indicates the presence of oxygen vacancies. The Ti 2*p* core peak does not show the presence of either $Ti^{3+}$ or $Ti^{2+}$. The large moment and high purity of the target clearly attributes the ferromagnetism to oxygen vacancies. The electronic structure of the reduced $TiO_2$ shows the occupied Ti 3*d* band and that the Fermi level falls on region of high density of states. This can lead to Stoner splitting of the band. The magnitude of the exchange splitting in the conduction band is estimated to be as large as 0.5 eV. Thus, the observed magnetism is due to the oxygen vacancy but more work has to be done to understand the distribution of defects and the coupling mechanism between the defects.

**References:**

1. Y.M. Gao, H.S. Shen, K. Dwight and A. Wold *Mater. Res. Bull.* **27**, 1023 (1992)

2. N. Negishi, K. Takeuchi and T. Ibusuki *Appl. Surf. Sci.* **121/122**, 417 (1997)

3. T. Dietl, H. Ohno, M. Matsukura , J. Cibert and D. Ferrand, *Science*, **287,** 1019 (2000)

4. J.W. Quilty J.W, A. Shibata, J.Y. Son, K.Takubo, T. Mizokawa, H.Toyosaki, T. Fukumura and M. Kawasaki, *Phys. Rev. Lett.* **96,** 0272021-4 (2006).

5 R. Janisch and N. A. Spaldin, *Phys. Rev. B* **73,** 035201 (2006)

6. K. Mamiya , T. Koide, A. Fujimori, H. Tokano, H. Manaka, A. Tanaka, H. Toyosaki, T. Fukumura and M. Kawasaki, *Appl. Phys. Lett.* **89**, 062506 (2006)

7. H. Munekata, H. Ohno, S. von Molnar, A. Segmuller, L. L. Chang, and L. Esaki, *Phys. Rev. Lett*. **63**, 1849 (1989)

8. H. Ohno *Science* **281**, 951 (1998).

9. M. Venketesan, C.B. Fitzgerald and J. M. D. Coey, *Nature* **430**, 630 (2005).

10. N. Hoa Hong, J. Sakai, N. Poirot and V. Brize, *Phys. Rev. B*. **73**, 132404 (2006).

11. S. D. Yoon, Y. Chen, A. Yang, T. L. Goodrich, X. Zuo, D. A. Arena, K. Ziemer, C. Vittoria and V. G. Harris, J. Phys.: Condens. Matter. 18, L355 (2006).

12. C. D. Pammaraju and S. Sanvito, *Phys. Rev. Lett*. **94**, 217205 (2005).

13. A.T. Paxton and L. Thien-Nga, *Phys. Rev. B* **57**, 1579 (1998).

14. J.B. Goodenough, *Phys. Rev. B* **5**, 2764 (1972).

15. E. Wahlström, E.K. Vestergaard, R. Schaub , A. Rønnau , M. Vestergaard, E. Lægsgaard, I. Stensgaard and F. Besenbacher, *Science* **303**, 511 (2004).

16. S. A. Chambers, S. M. Heald, R. F. C. Farrow, J. U Thiele, R. F. Marks, M. F. Toney and Chattopadhyay, *Appl. Phys.Lett*. **79**, 3467 (2001).

17. M. Venketesan, C.B. Fitzgerald, J. G. Lunney and J. M. D. Coey, *Phys. Rev. Lett* **93**, 177206 (2004).

18. N. H. Hong, N. Poirot and J. Sakai, *Appl. Phys.Lett.* **89**, 042503 (2006).

19. D. W. Abraham, M. M. Frank and S. Guha, *Appl. Phys. Lett*. **87**, 252502 (2005).

20. B. Holmberg, *Acta Chem. Scand.* **16**, 1245 (1962).

21. K. H. Kim, K. J. Lee, D. J. Kim, H. J. Kim, Y. E. Ihm, C G. Kim, S. H. Yoo and C. S. Kim , *Appl. Phys. Lett*. **82**, 4755 (2003).

22. G. Tyuliev and S. Angelov, *Appl. Surf. Sci*, **32**, 381 (1988).

23. M. Naeem, S. K. Hasanain, M. Kobayashi, Y. Ishida, A. Fujimori, S. Buzby and S. I. Shah, *Nanotechnol.* **17**, 2675 (2006).

24. M. Murata, K. Wakino and S. Ikeda, *J. Elect. Spectros*. **6**, 459 (1975).

25. P. M. Kumar, Badrinarayanan and M. Sastry , *Thin Solid Films* **358**, 122 (2000).

26. K. E. Smith, J.L. Mackay and V. E. Henrich, *Phys. Rev. B*, **35**, 5822 (1987)

27. V. E. Henrich and P. A. Cox, *The Surface Science of Metal Oxides*, 1st ed. Cambridge University Press, Cambridge (1994).

28. H. Toyosaki, T. Fukumura, Y. Yamada, K. Nakajima, T. Chikyow, T. Hasegawa, H. Koinuma and M. Kawasaki, *Nat. Mater.* **3**, 221 (2004)

**Figure Captions**

Figure-1. XRD patterns for pure $TiO_2$ (undoped), obtained for the samples grown under different oxygen partial pressures

Figure-2 (a). Magnetization versus magnetic field at 298 K for pure $TiO_2$ films grown under different oxygen partial pressure. The inset shows magnetization of pure target moment (scale arbitrary). (b) Magnetization versus temperature for 0.02 mTorr sample

Figure-3. XPS survey spectra for $TiO_2$ film. The peaks in the box are Auger peaks of $O_2$ and Ti.

Figure 4 (a) . XPS spectra of O 1*s* core level for samples prepared at 0.02 mTorr and 50 mTorr. (b) Relative area of the HBEC peak as a function of oxygen partial pressure during deposition.

Figure 5. Ti 2*p* XPS core-level spectra for samples prepared at 0.02 mTorr and 50 mTorr O partial pressure/

Figure 6. Valence band XPS spectra sample prepared at 0.02 mTorr (reduced) and 50 mTorr $TiO_2$ film. The inset shows the magnified region of the Ti 3*d* band

Figure 7. O 1s and Ti 2p XPS core level peaks for different oxygen partial pressure showing the chemical potential shift.

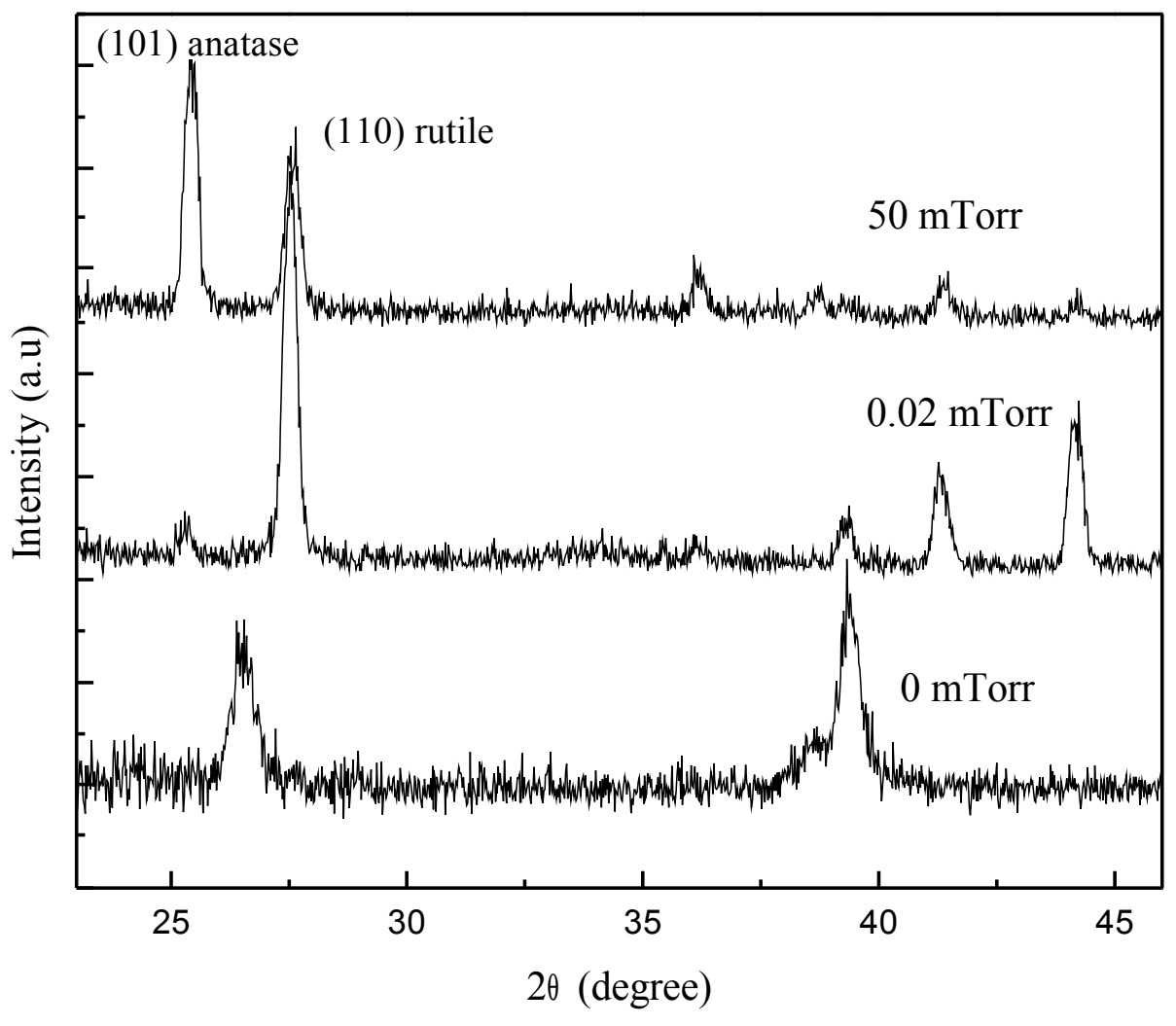

Figure 1. Abdul K Rumaiz et al

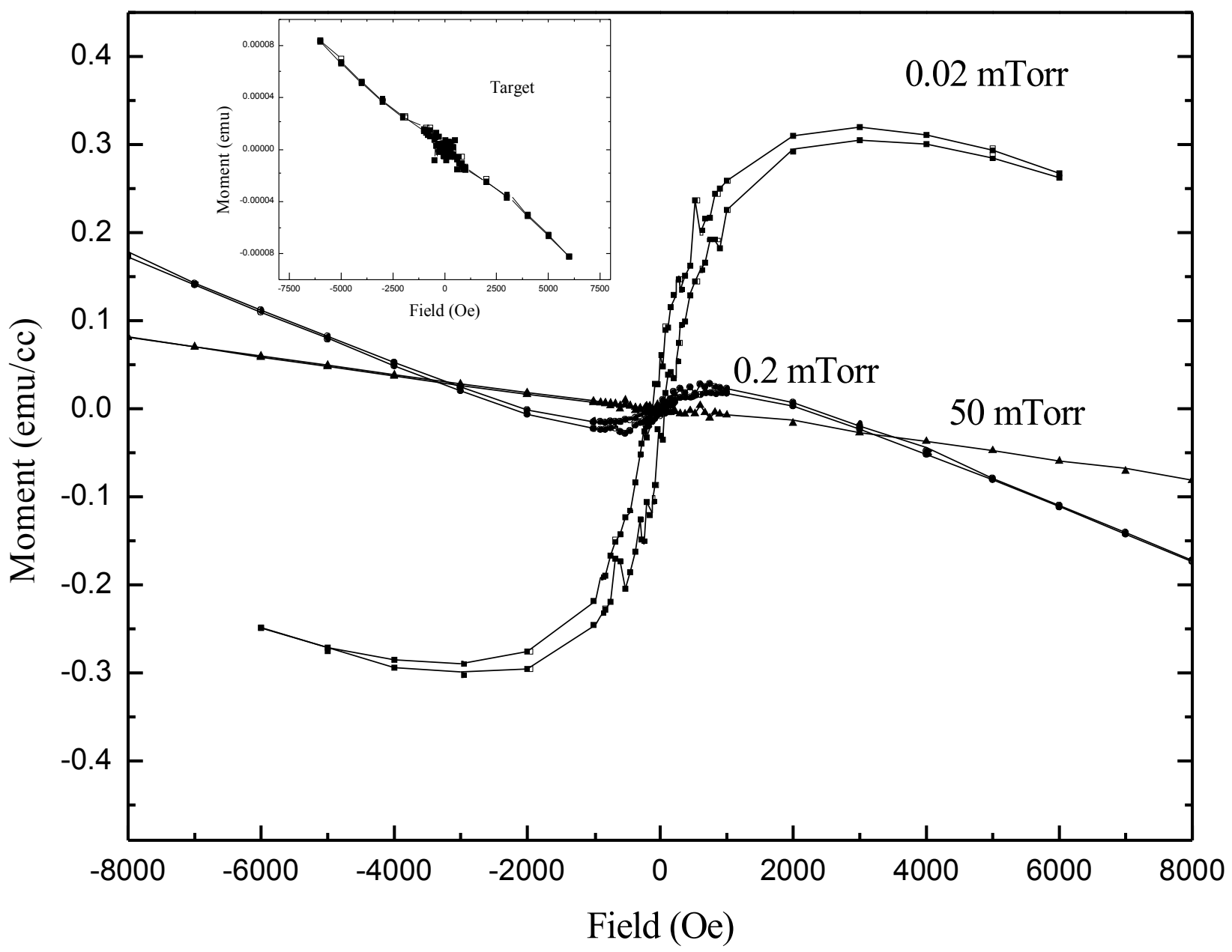

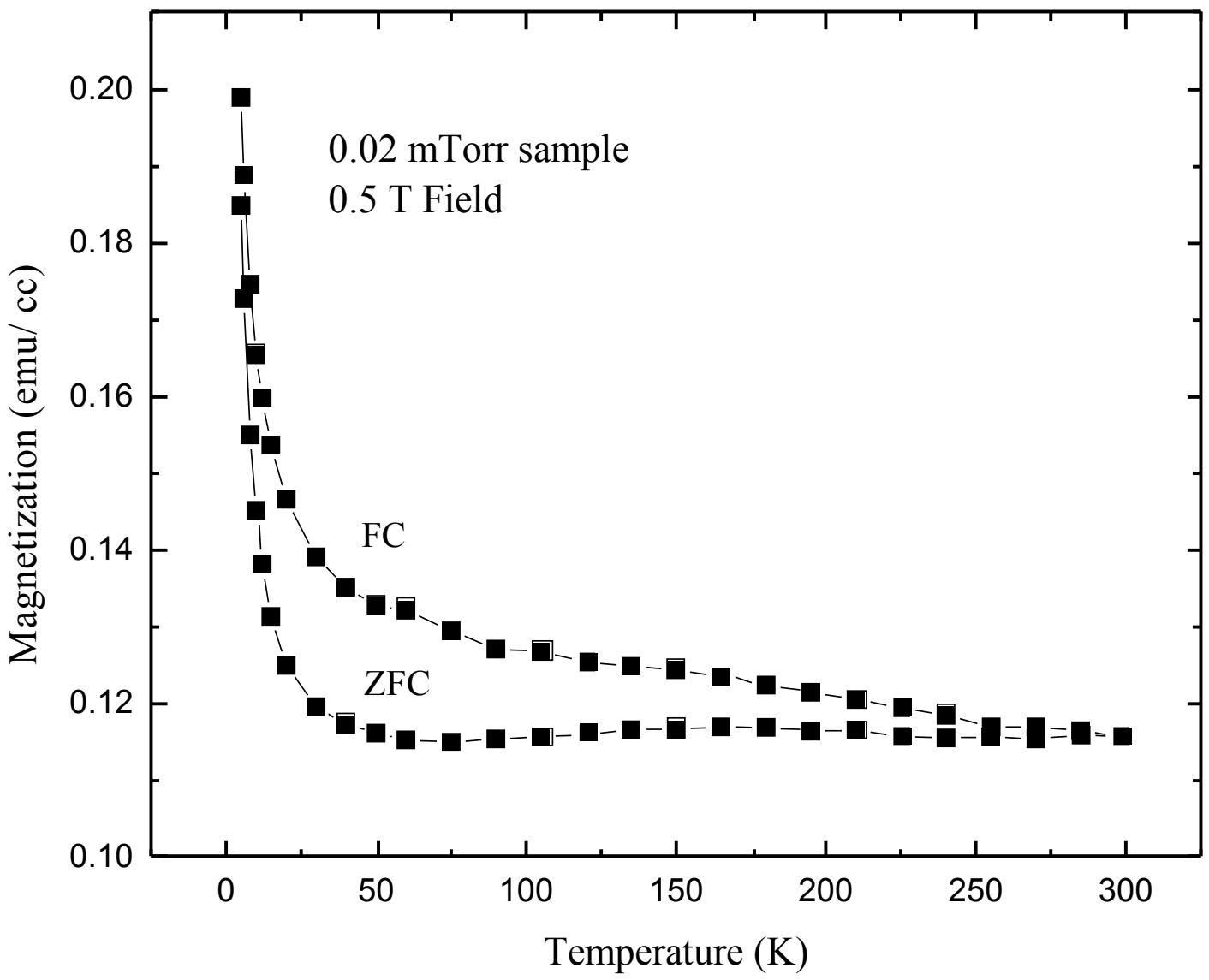

Figure 2 a & b

. Abdul K Rumaiz et al

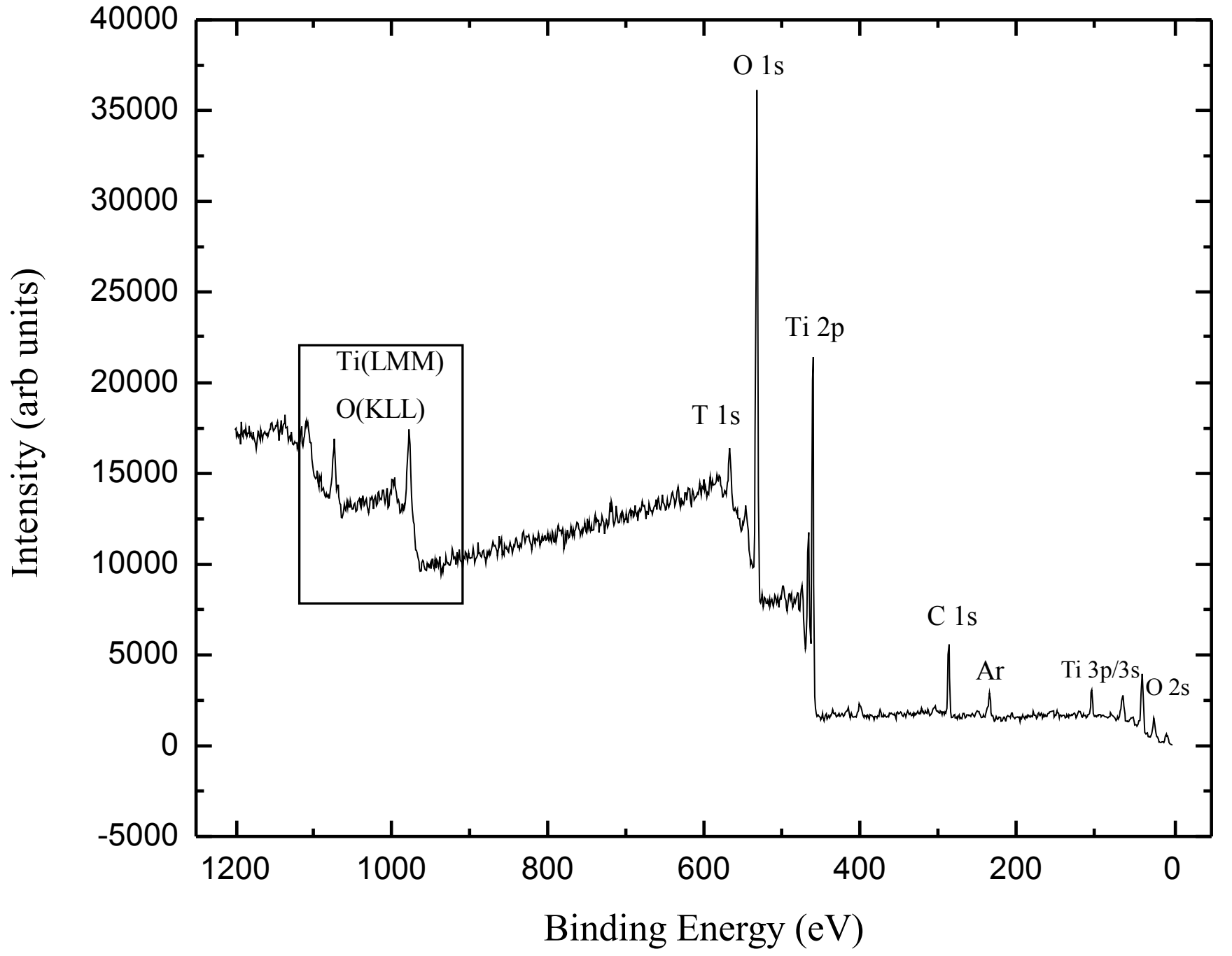

Figure 3. Abdul K Rumaiz et al

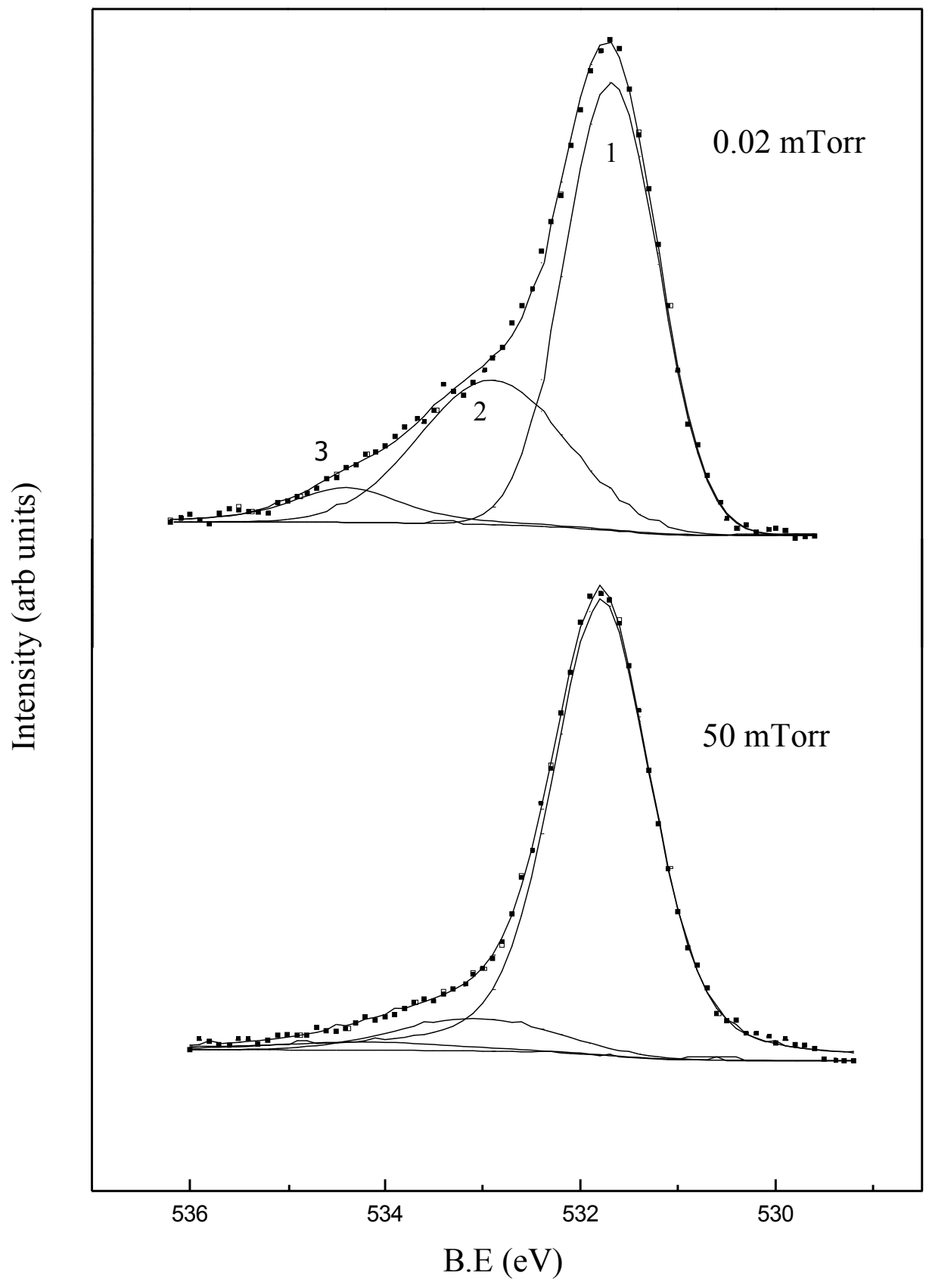

Figure 4 a. Abdul K Rumaiz et al

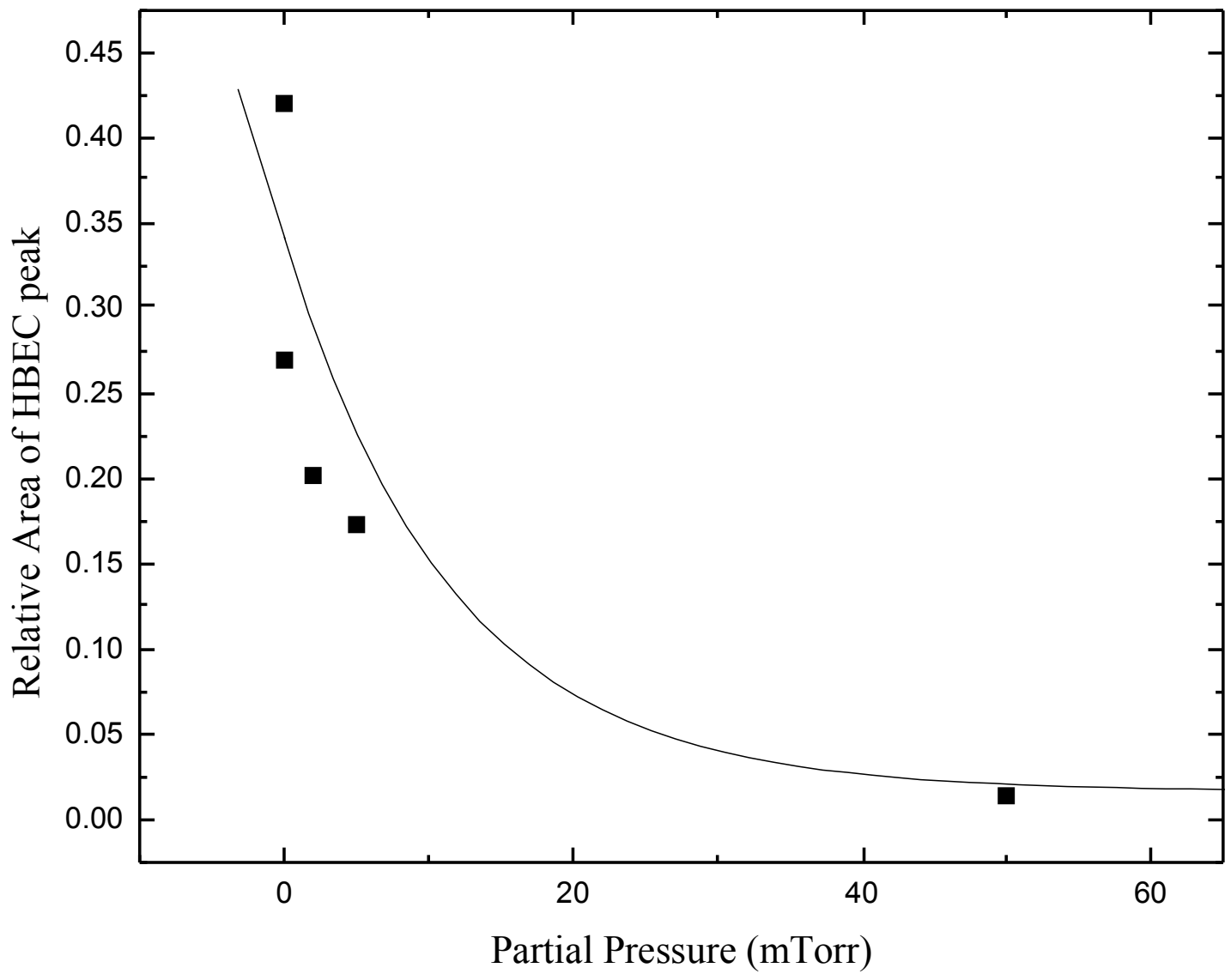

Figure 4 b. Abdul K Rumaiz et al

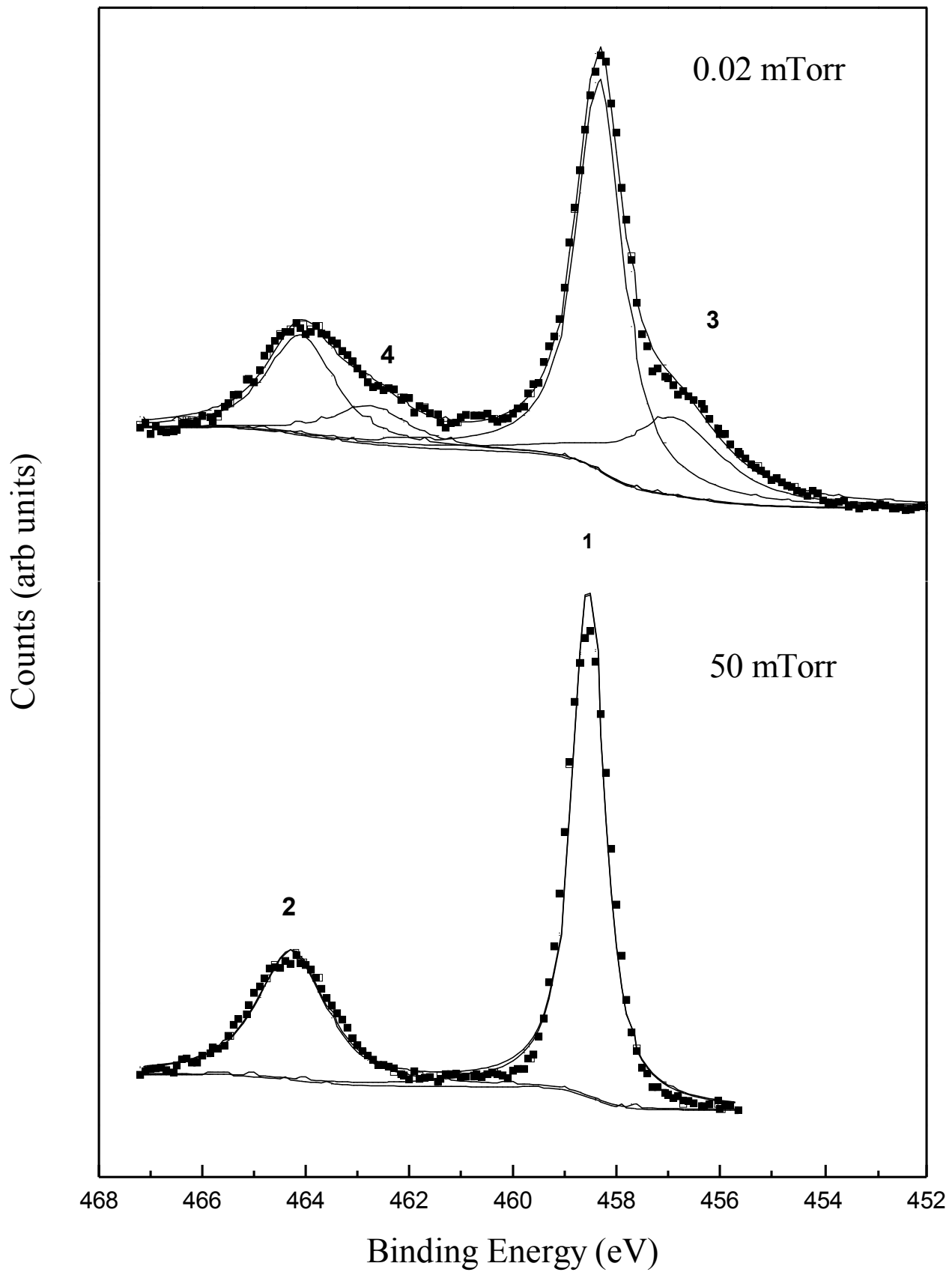

Figure 5 Abdul K Rumaiz et al

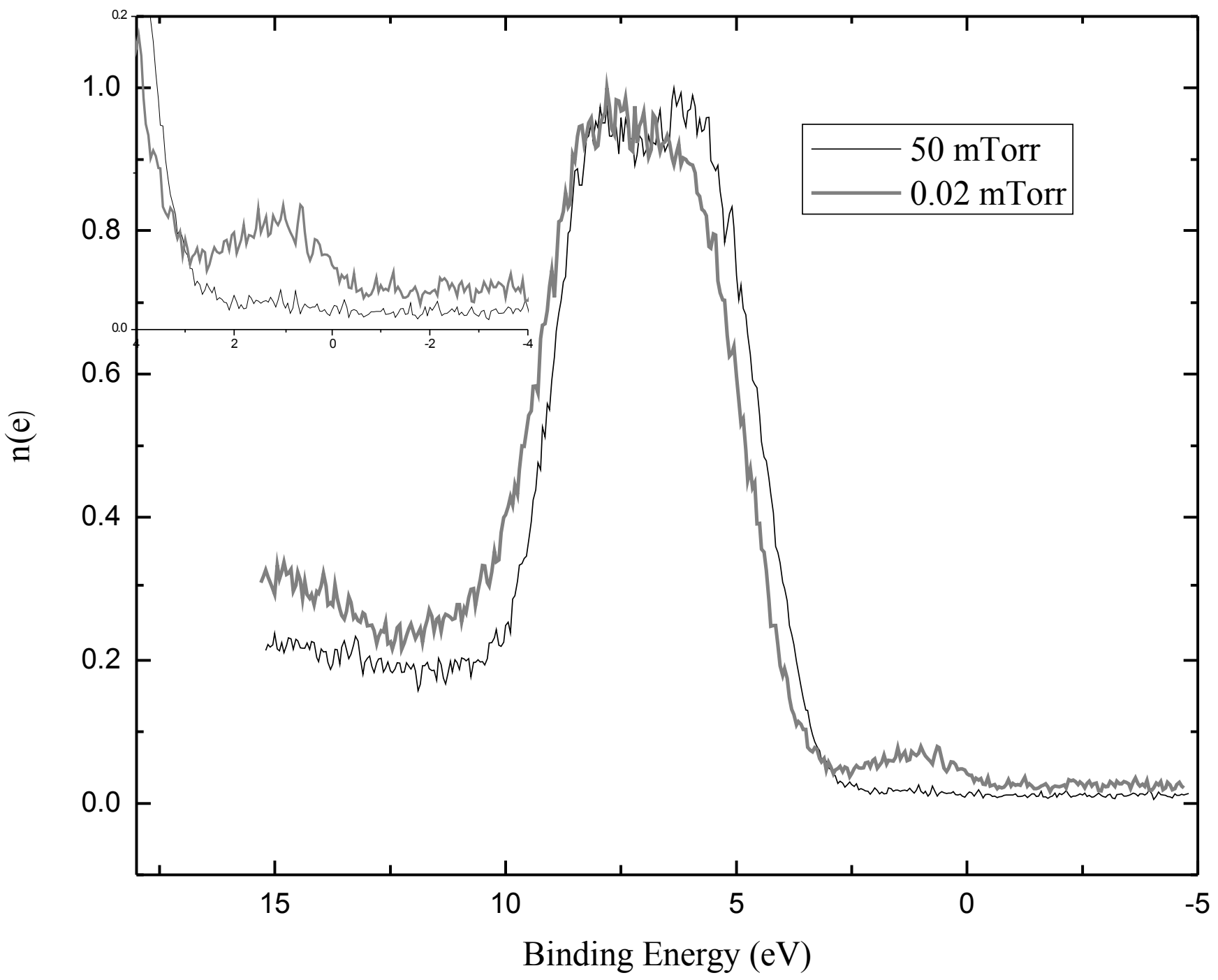

Figure 6. Abdul K Rumaiz et al

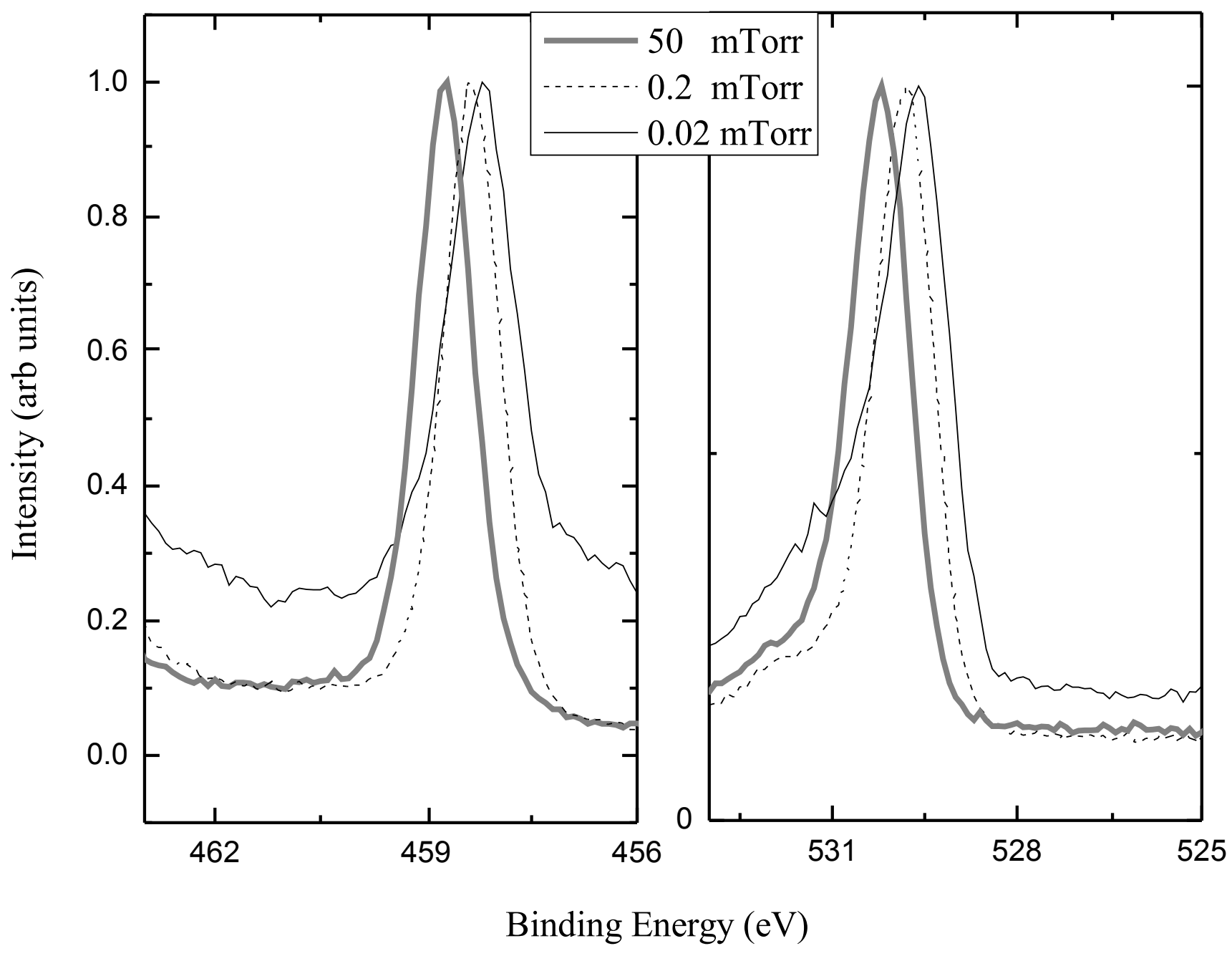

Figure 7. Abdul K Rumaiz et al